\documentclass[a4paper]{article}
\usepackage{color,url}
\usepackage{INTERSPEECH2021}
\usepackage{multirow}
\title{The ZevoMOS entry to VoiceMOS Challenge 2022}
\name{Adriana STAN}
\address{
  Communications Department, \\Technical University of Cluj-Napoca, Romania}
\email{adriana.stan@com.utcluj.ro}

\begin{document}

\maketitle
\begin{abstract}
  This paper introduces the ZevoMOS entry to the main track of the VoiceMOS Challenge 2022.
  The ZevoMOS submission is based on a two-step finetuning of pretrained self-supervised learning (SSL) speech models. The first step uses a task of classifying natural versus synthetic speech, while the second step's task is to predict the MOS scores associated with each training sample. The results of the finetuning process are then combined with the confidence scores extracted from an automatic speech recognition model, as well as the raw embeddings of the training samples obtained from a wav2vec SSL speech model. 
  
  The team id assigned to the ZevoMOS system within the VoiceMOS Challenge is T01. The submission was placed on the 14th place with respect to the system-level SRCC, and on the 9th place with respect to the utterance-level MSE. The paper also introduces additional evaluations of the intermediate results. 
  
\end{abstract}

\section{Introduction}

The evaluation of text-to-speech (TTS) synthesis systems still heavily relies on performing subjective listening tests. In these tests, several factors can affect the final judgement of the listeners with respect to a audio sample, such as: \texttt{exposure}--did the listener ever listen to synthetic samples before; \texttt{context}--what other audio systems and audio samples are present in the same listening test;  or \texttt{failure-to-justify}--meaning that the listener cannot pinpoint exactly why a sample is bad, it is just that it does not sound natural. 

This means that, aside from the additional and lengthy effort of preparing and conducting listening tests, their results are in some cases inconclusive or error prone. Another problem with the listening tests is that the listener only rates the samples at an utterance level, commonly on a scale of 1 to 5. This discrete scale, and high level evaluation does not provide sufficient information for the TTS developer on how to improve the system. And also, in the next iterations of the TTS development, the listening tests have to be conducted again. 

The most common sections of a listening test pertain to naturalness, speaker similarity and intelligibility. In recent years, the latter two sections have seen some alternative objective measures being applied to their evaluation. For the speaker similarity, large speaker-verification networks can be involved \cite{vgg}. The average embeddings, or equal error rates computed on the synthesised samples and the natural samples are beginning to show acceptability in the TTS community. 
On the other hand, for the intelligibility, highly-accurate automatic speech recognition systems and their word error rates (WER) can enable a measure of how well the TTS system pronounced the input text \cite{asr-tts}. 

As a result, the last elusive objective measure of TTS quality lies within the most complex one, the naturalness. This is also the hardest measure to implement, as even natural samples are sometimes rated unnatural when minimum levels of noise and reverberation are present in the sample. Take for example the listening tests performed for Tacotron2~\cite{tacotron2} and FastPitch~\cite{fastpitch} where there is an 0.5 difference between the average scoring of the natural samples. 
Yet, finding an objective measure of audio quality is not only useful in the final evaluation of TTS or voice conversion systems, but also, if adequate, could be used as loss measures in the training process of such systems. Nowadays, the loss of DNN-based speech systems uses L1 or L2 losses, which do not truly correlate with the perceived quality or naturalness of the output speech.

There are several works which attempt to solve the objective evaluation of the synthesised samples' issue.  
For example MOSNet~\cite{mosnet} uses three DNN architectures, CNN, BLSTM, and CNNBLSTM to predict the MOS scores of the Voice Conversion Challenge 2018 using frame- and utterance-level predictions of the MOS scores. MBNet~\cite{mbnet} introduces an additional sub-network which accounts for the individual listeners' evaluations of each utterance and aggregates these scores into the final MOS prediction. A similar approach is taken by LDNet~\cite{DBLP:journals/corr/abs-2110-09103} where individual listener scores, as well as utterance-averaged scores are predicted. LDNet also evaluates multiple NN architectures, such as convolutional 2D networks, MobileNetV2~\cite{mobilenetv2} and MobileNetV3~\cite{mobilenetv3}. The training dataset is also the Voice Conversion Challenge 2018 listening test's results. 
\cite{williams} looks into different embeddings of the audio samples, such as deep spectrum, x-vectors and acoustic embeddings in conjunction with the MOSNet, and with both speaker-level and system-level aggregation of the MOS scores. 
One interesting study is that of~\cite{tseng2021utilizing} where the authors look into several self-supervised models, such as wav2vec~\cite{wav2vec}, CPC~\cite{cpc}, APC~\cite{apc} and Tera~\cite{tera}, and finetune them in an end-to-end manner to perform the naturalness score predictions on the Voice Conversion Challenge 2018 data. 
An extensive analysis of the MOS predictions is performed by the authors of~\cite{generalisation}, where they compare several previously published MOS prediction networks, and pretrained self-supervised models in a task of in-domain, as well as out-of-domain MOS prediction task. 

\begin{figure*}[!ht]
  \centering
  \includegraphics[width=0.7\linewidth]{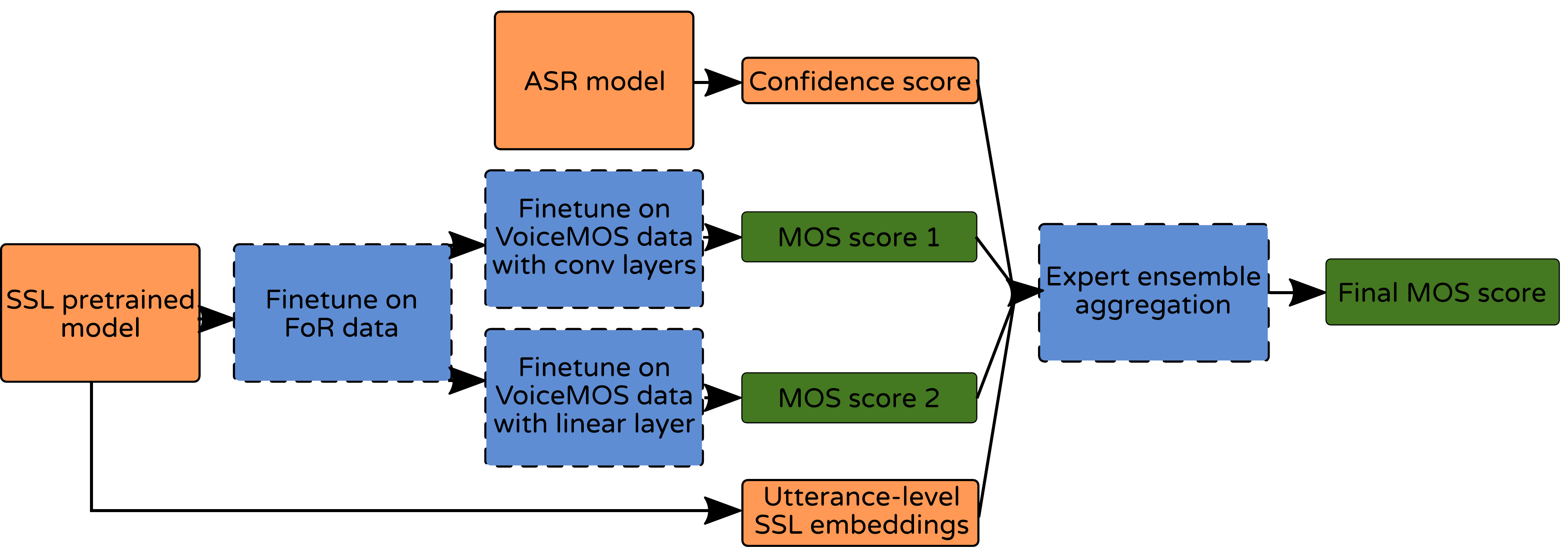}
  \caption{Overview of the ZevoMOS system's architecture. The orange boxes mark the pretrained modules, and the blue boxes mark the modules trained for the VoiceMOS Challenge predictions.}
  \label{fig:flow}
\end{figure*}

In this work, we introduce the results of our MOS prediction system, named ZevoMOS, entered into the 2022 VoiceMOS Challenge~\cite{voicemos}. The system builds on top of the  self-supervised speech learning (SSL) representations finetuned in a natural versus synthetic speech classification task, and aggregates the SSL-based predictions with the confidence scores of an ASR system. The paper is organised as follows: Section \ref{sec-syst} introduces the general structure of the ZevoMOS system and its additional inputs. Section \ref{sec-res} describes the data, initial and VoiceMOS challenge results for our system, while Section~\ref{sec-conc} discusses the results and draws the conclusions. 

\section{ZevoMOS system description}
\label{sec-syst}

The development of the ZevoMOS prediction system starts from the baseline networks provided by the VoiceMOS Challenge organisers. Three different systems were provided, and we selected the one described in~\cite{generalisation} and its implementation.\footnote{\url{https://github.com/nii-yamagishilab/mos-finetune-ssl}}
In this baseline, the MOS predictions are learned by adding a final linear layer to a SSL model and finetuning the model in an end-to-end manner. The following subsections present the changes made to this baseline model, and the additional input and score aggregation method used in ZevoMOS.

\subsection{Self-supervised learning speech models}

The experiments presented in \cite{generalisation} use the wav2vec pretrained models available in the Fairseq~\cite{fairseq} repository.\footnote{\url{https://github.com/pytorch/fairseq/tree/main/examples/wav2vec}} However, these models are trained on natural speech alone. We therefore started out by finetuning the model on the FoR dataset~\cite{8906599} in a natural versus synthetic speech classification task. The FoR dataset contains around 198,000 English utterances from some of the latest commercial deep-learning speech synthesizers (i.e. Amazon, Google and Baidu), as well as natural speech from the Arctic, LJSpeech and VoxForge datasets. Such that none of the synthetic samples overlap with the systems used to collect the VoiceMOS data. This was one of the rules of the challenge, which stated that the data from the ESP-Net TTS, Blizzard and Voice Conversion challenges should not be used to train the prediction models. However, the FoR dataset contains only high-quality synthetic speech samples, which are not as well correlated to, for example, the HMM-based TTS systems that are present in the Blizzard Challenge evaluations. Still, we considered that having the SSL models exposed to a limited set of artefacts which appear in the synthesised samples was important, and the numeric results support our initial hypothesis. 
As a result, the wav2vec hidden representations of the FoR samples averaged over the entire utterance were passed through a final linear layer to output a natural or synthetic binary classification label. The information from the classification task was enabled to flow back into the wav2vec network and update its weights. A similar finetuning step was performed for the Decoar 2.0~\cite{decoar2} pretrained model available in the S3PRL repository.\footnote{\url{https://github.com/s3prl/s3prl}} The models were only trained for a few epochs ($<10$) such that the bias of the classification task does not affect the learned audio embeddings.

The resulting wav2vec and Decoar 2.0 models were then finetuned on the VoiceMOS training data. The models' hidden representations were passed through either an average pooling layer followed by a linear layer, or a sequence of two 2D convolution layers, followed by three linear layers. The task of this finetuning step was to predict the utterance-level MOS scores. The models were trained with an early stopping criterion which looked into the MSE values computed over the dev set.

\subsection{ASR-based confidence scores}

Part of the evaluation of the naturalness of a synthetic sample pertains to its intelligibility. The intelligibility is also commonly evaluated in subjective listening tests. Yet, recent studies showed that high-quality ASR systems can give a good indication of the intelligibility of synthetic samples, without the definite need for a subjective evaluation of this measure~\cite{asr-tts}. 
The intelligibility of synthetic samples is measured as the word error rate between the input/target text and the transcriptions of that sample performed by either the listeners, or an ASR system. 
However, in the VoiceMOS challenge, the transcripts of the training samples were not available. 
We therefore relied on the confidence scores exhibited by the 
Alpha Cephei generic US English model\footnote{\url{https://alphacephei.com/vosk/models}}. The self-reported WER on the LibriSpeech dataset is 5.69\%. The model selection was made also because the Alpha Cephei ASR outputs word-level confidence scores. The average of these word-level confidence scores across the entire utterance was used as additional input into the final MOS predictions made by the ZevoMOS system. It is worth mentioning here that an initial confidence score estimation was performed with the Google Speech-to-Text API.\footnote{\url{https://cloud.google.com/speech-to-text}} Yet, for the very poor-quality samples, the API failed to output any transcript or confidence measure. This left around 100 samples in both the training and the dev sets without this measure, and we therefore dropped the use of this API.

\subsection{Expert ensemble aggregation}

The individual predictions of the finetuned SSL models or the ASR-based confidence scores exhibited sub-par results on the dev data. We present these individual performances in the Results section. 
However, we noticed that some of the predictions made by individual components were complementary. Such that we resorted to aggregate the individual predictions in a final expert ensemble module. The ensemble included the predictions of the SSL finetuned models, the ASR confidence scores, and augmented with the baseline wav2vec hidden representations of the audio samples. The aggregation was performed with either a LightGBM~\cite{NIPS2017_6449f44a} method, or a three linear layer neural net. Both structures used equal input weights for each component.
A diagram of the final ZevoMOS system is shown in Figure~\ref{fig:flow}, and the source code to train the system, as well as the pretrained models are available at the following link: \url{https://github.com/adrianastan/ZevoMOS/}.

\section{Evaluation}
\label{sec-res}

\subsection{The VoiceMOS Challenge 2022 Speech Dataset}
The VoiceMOS Challenge 2022 includes two tracks: a \texttt{MAIN} track in which all the samples were evaluated within the same listening test by the same listeners, and an out-of-domain track (\texttt{OOD}), in which the samples were evaluated in a separate listening test. Also, the main track comprises English samples, while the OOD track includes Chinese synthetic samples. The ZevoMOS entry made submissions only to the \texttt{MAIN} track of the challenge.  

The VoiceMOS Challenge 2022 data for the \texttt{MAIN} track is composed of a subset of utterances gathered from the past Blizzard and Voice Conversion Challenges. These samples were reevaluated in a new large-scale listening test~\cite{Cooper2021HowDV}. The listening test contained samples from 187 different systems covering various speech technologies, from unit selection to end-to-end neural systems.
The data is sampled at 16kHz and was released as separate training, dev and test sets. The training set comprises 4974 samples from 175 systems. The dev set contains 1066 samples from 181 systems. 6 systems (amounting to 222 samples) from the dev set were not present within the training data, however their samples were evaluated in the same large-scale listening test. These two datasets were available in the training phase of the challenge, and could be used to evaluate the initial models in the challenge's platform.  
A separate dataset, the test set, containing 1066 samples from 187 systems was used to perform the final evaluation of the submissions. The target MOS scores of the test set utterances were made available only after the completion of the evaluation phase. 12 systems present in the test set, with a total of 234 utterances were not present in the training set, however 6 of these systems were present in the dev set. But during the test phase, the dev set was not supposed to be used as additional training data. 

Due to compute power limitations, we had to restrict the length of the models' input files to 6 seconds in both the training and testing procedures. This, of course, might have led to sub-optimal predictions, as some artefacts might have only occurred later in the audio sample. The data released in the VoiceMOS Challenge also included listener information and individual listener ratings. Each sample was rated by 8 listeners, but the ZevoMOS entry did not take this information into account, and only used the sample-averaged MOS scores.  

\subsection{Initial evaluation}


\begin{table}[th!]
  \caption{MOS scores' prediction results on the \textbf{dev} dataset for the individual SSL models. The best results in each column are marked in boldface.}
  \label{tab:ssl-val}
  \centering
  \begin{tabular}{ l| c|c|c|c}
    \toprule
     & \multicolumn{2}{c}{\textbf{System}} & \multicolumn{2}{|c}{\textbf{Utterance}} \\
     \cmidrule{2-5}
    \textbf{Model} & \textbf{SRCC} & \textbf{MSE} & \textbf{SRCC}  & \textbf{MSE} \\
    \midrule
      w2v baseline   & 0.919 & 0.119 & 0.856 & 0.241 \\
      w2v-FoR-linear & 0.927 & \textbf{0.110} & 0.865& \textbf{0.253}  \\
      w2v-FoR-conv   & 0.932 & 0.310 & \textbf{0.872}& 0.482\\
      \midrule
      decoar2 baseline   & 0.890 & 0.214& 0.824 & 0.352  \\
      decoar2-FoR-linear & 0.927 & 0.120& 0.856 & 0.278  \\
      decoar2-FoR-conv   & \textbf{0.935} & 0.264& 0.863 & 0.350   \\
    \bottomrule
  \end{tabular}
\end{table}

\begin{figure}[h!]
  \centering
  \includegraphics[width=\columnwidth]{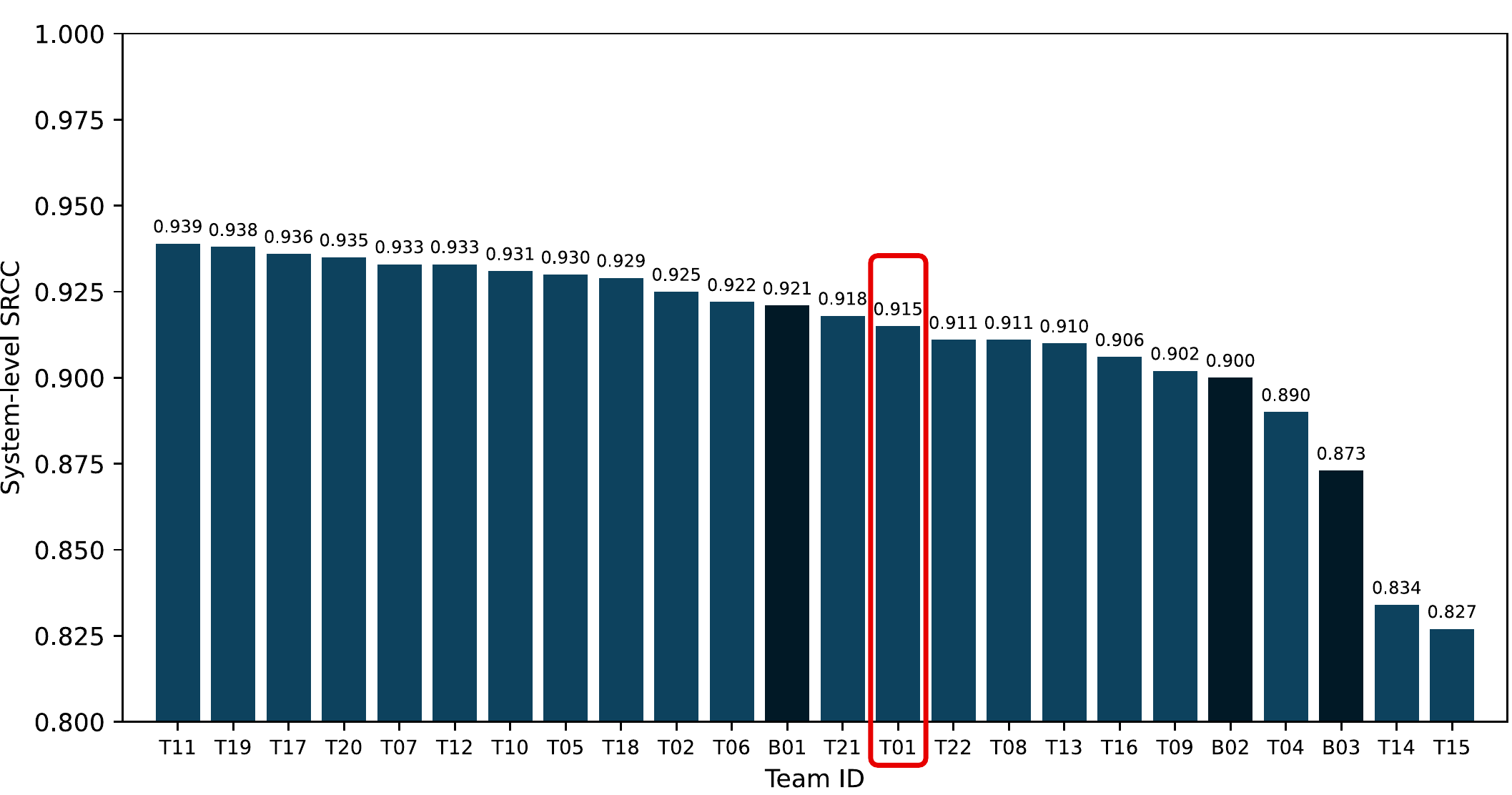}
    \centering{(a)}
  \centering
  \includegraphics[width=\linewidth]{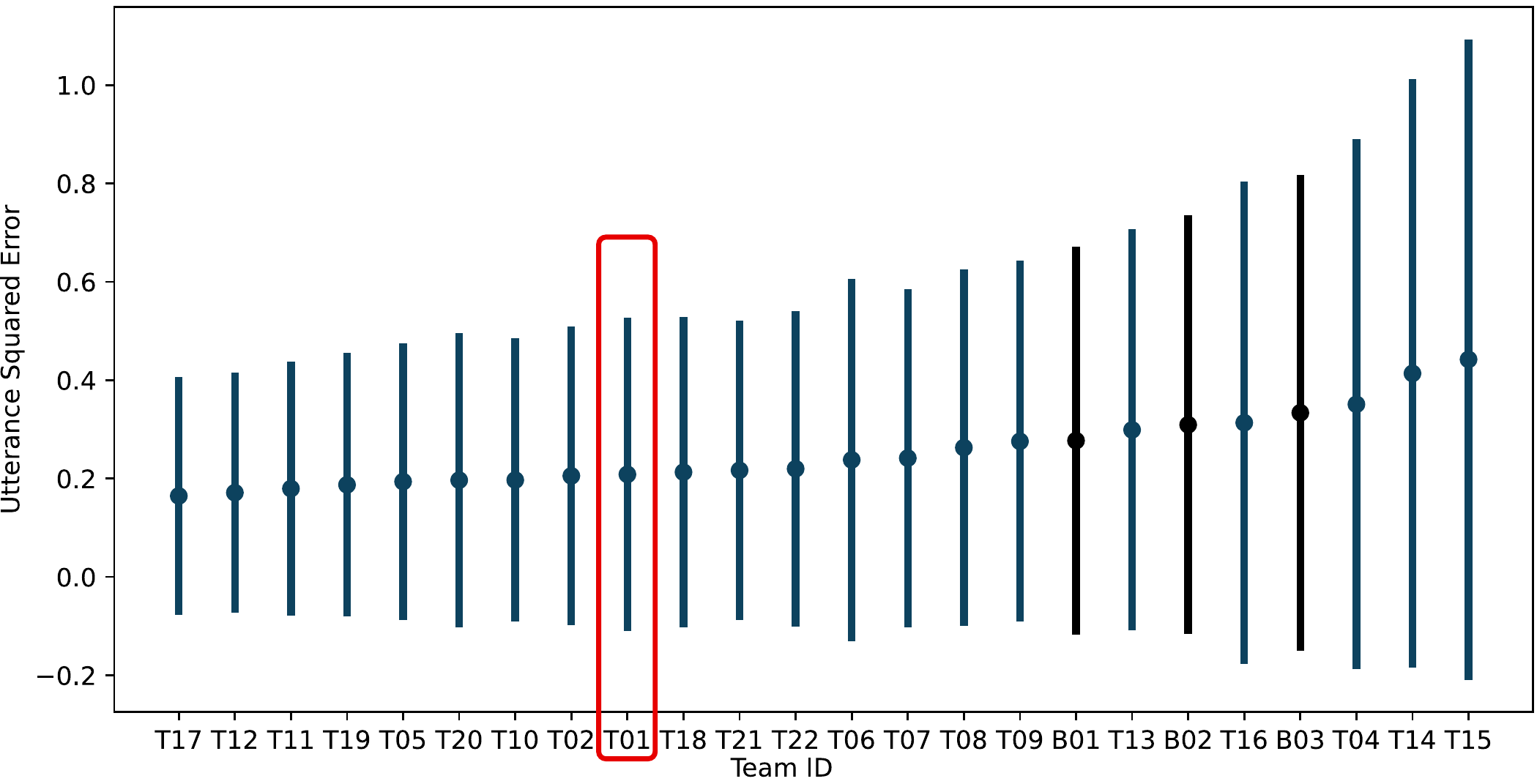}
  \centering{(b)}
  \caption{VoiceMOS main track \textbf{(a) system-level SRCC rankings} and \textbf{(b) utterance-level MSE rankings}. The ZevoMOS system is under team id \texttt{T01}. The darker shades represent the baseline systems.}
  \label{fig:srcc}
\end{figure}

Prior to submitting the ZevoMOS score predictions to the VoiceMOS Challenge scoreboard, a series of initial evaluations were performed. The objective evaluation of the VoiceMOS Challenge 2022 uses the Spearman Rank Correlation Coefficient (SRCC) and the Mean Squared Error (MSE) of the MOS scores for individual utterances, as well as the averaged scores across a system's samples. Therefore, there are 4 objective measures to analyse: \emph{utterance-level SRCC and MSE}, and \emph{system-level SRCC and MSE}. 

The first thing we wanted to look into were the system-level correlations between the VoiceMOS data subsets. There is an $0.917$ SRCC between the training and dev sets and a $0.905$ SRCC between the training and the test sets. In theory, without any additional information, and assuming similar data distributions across the datasets, these would be the top-most values that a system trained only on this data could attain. We also noticed that for part of the systems available only in the dev or test sets, there was only one sample available. In this case the system-level mean MOS value is equal to the rating of that particular sample.

\begin{table*}[t!]
  \caption{Results on the \textbf{dev} and \textbf{test} data for various expert ensemble input components. The components' numbering refer to: [A] ASR confidence scores, [B] wav2vec baseline embeddings, [C] wav2vec-FoR-linear, and [D] wav2-FoR-conv. The ZevoMOS submission includes all four components, but uses a neural network for ensembling, instead of LightGBM.}
  \label{tab:abl-val}
  \centering
  \begin{tabular}{c|c|c|c| |c|c|c|c| |c|c|c|c}
    \toprule
      \multicolumn{4}{c}{\textbf{Components}} & \multicolumn{4}{c}{\textbf{Dev dataset}} & \multicolumn{4}{||c}{\textbf{Test dataset}} \\
    
     \multicolumn{4}{c}{\textbf{}}  & \multicolumn{2}{c}{\textbf{System}} & \multicolumn{2}{|c}{\textbf{Utterance}}& \multicolumn{2}{||c}{\textbf{System}} & \multicolumn{2}{|c}{\textbf{Utterance}} \\
     
     \cmidrule{5-12}
    [A]&[B]&[C]&[D]& SRCC & MSE & SRCC & MSE& SRCC & MSE & SRCC & MSE \\
    \midrule
      \checkmark &  &  &  & 0.522 & 0.557 & 0.296 & 0.767 & 0.482 & 0.539 & 0.394 & 0.717 \\
       & \checkmark &  &  & 0.905 & 0.130 & 0.825 & 0.261 & 0.894 & 0.150 & 0.835  & 0.260 \\
      \checkmark & \checkmark &  &  & 0.906 & 0.130&  0.825 & 0.261  & 0.907  &  0.140& 0.842 & 0.250 \\
      
     \checkmark & &  &  \checkmark  & 0.929 & 0.096 & 0.865 & 0.208 & 0.912 & 0.126 & 0.868 & 0.214 \\
      \checkmark& & \checkmark &    &0.931 & 0.090 & 0.860 & 0.216 & 0.912& 0.127 & 0.853 & 0.248 \\
     & \checkmark & \checkmark &    & 0.935 &0.091 & 0.873 & 0.198& 0.911& 0.124 & 0.870 & 0.219 \\
     & & \checkmark &  \checkmark   & 0.941 & 0.083 & 0.876 & 0.194 & 0.913 & 0.123 & 0.869 & 0.220 \\
    & \checkmark& \checkmark &  \checkmark    & 0.941 & 0.078 & 0.883 & 0.181 & 0.916& 0.120 & 0.874 & 0.210 \\
    \checkmark & \checkmark& \checkmark &  \checkmark      & 0.942 & 0.070 & 0.883 & 0.180 & 0.917 & 0.119 & 0.873 & 0.211\\
    \midrule
    \multicolumn{4}{c||}{\textbf{ZevoMOS submission} }& 0.951 & 0.067 &0.883 & 0.182 &0.915 & 0.122 &0.878 & 0.208\\
      
    \bottomrule
  \end{tabular}
\end{table*}

We then proceeded to evaluate the intermediate models obtained in the training process of the ZevoMOS system.
The results of this evaluation over the dev dataset are shown in Table~\ref{tab:ssl-val}. The models pertain to the following definitions: \texttt{w2v baseline} - the pretrained wav2vec model finetuned on the MOS scores using a single linear output layer; \texttt{w2v-FoR-linear} - the wav2vec model finetuned on the FoR dataset and finetuned again to predict the MOS scores using a single linear output layer; \texttt{w2vFoR-conv} - the wav2vec model finetuned on the FoR dataset and finetuned again to predict the MOS scores using the sequence of 2D convolution layers followed by the three linear output layers. The same notations apply for the Decoar 2.0 model.\footnote{We should note here that although we started from the \texttt{B01}  baseline provided by the challenge organisers, we retrained the MOS prediction model and also used the large version of the wav2vec. Therefore, our results for this baseline differ slightly.}
It can be noticed that the simple finetuning of SSL models using the natural versus synthetic classification task improves the MOS score predictions' accuracy for both the wav2vec and Decoar 2.0 models. The choice of the sequence of convolution plus linear output layers also add a marginal improvement to the SRCC measures. However, compared to the single linear output layer, the convolution sequence drastically affects the MSE values. 
Overall, the wav2vec model performs better using the baseline model, yet after the finetuning processes, the results of wav2vec and Decoar 2.0 models are very similar. We eventually selected the wav2vec model for the final submission, as we focused more on the utterance-level SRCC measure.  

Starting from these initial results, we evaluated the different expert ensembles. The two structures, the LightGBM and linear NN aggregation methods exhibited very similar results, with only some light improvements on the NN side. Table~\ref{tab:abl-val} presents the results of various component combinations using the LightGBM method. However, the ZevoMOS submission (marked as such in Table~\ref{tab:abl-val}) used the NN-based ensemble, due to its slightly better results. It can be noticed that, although no major improvement is obtained by any of the individual components, their combined knowledge is incrementally beneficial.

\subsection{VoiceMOS Challenge evaluation}

Figure~\ref{fig:srcc} shows the results of all the submissions made to the challenge. The ZevoMOS system is under the \texttt{T01} team id. With respect to utterance-level MSE, ZevoMOS was ranked $9th$, and on the $14th$ place with respect to the system-level SRCC. The best performing systems obtained a system SRCC of $0.939$ and an utterance MSE of $0.165$. The results do not pertain to the same submission. It is interesting to notice that among the first 10 ranked systems w.r.t. system-level SRCC there is only a $0.01$ difference. The same variation applies for the first 15 ranked systems in terms of utterance-level MSE. 

The complete results of the ZevoMOS entry for both the dev and the test datasets are shown in Table~\ref{tab:abl-val}. In the training phase, ZevoMOS was placed on the $6th$ place w.r.t. system-level SRCC and on the $5th$ place w.r.t. utterance-level MSE.\footnote{\url{https://codalab.lisn.upsaclay.fr/competitions/695#results}}


\section{Discussions and Conclusions}
\label{sec-conc}

This paper introduced the ZevoMOS entry to the VoiceMOS Challenge 2022. The system uses an expert ensemble to aggregate individual MOS predictions made by self-supervised learning models. The SSL models were previously finetuned in a classification task of synthesised versus natural speech prediction. The expert ensemble also includes the confidence scores of a high-quality ASR system, as well as the baseline SSL model embeddings of the training/testing audio samples. The results placed the ZevoMOS system on the 14th place in terms of system-level SRCC and on the 9th place in terms of utterance-level MSE. 

Although the results of the ZevoMOS submission are behind the best performing systems, it still supports the idea that models pretrained on large amounts of natural speech samples are informative in terms of estimating the quality of a synthetic sample. And therefore, their study w.r.t. naturalness estimation should be further explored. It is true, however, that an ideal MOS predictor would be a simple signal-based differentiable measure which would ensure its use within the training process of synthetic speech generation architectures. As future work, we plan to investigate such signal-based measures which might be indicative of quality degradations of speech without a reference signal, but also to investigate other model architectures pretrained on large amounts of natural speech data, and which might have learned abstract features pertaining to the naturalness of the output signal.

\textbf{Acknowledgements.}
This work was funded by Zevo Technology SRL. The author would like to thank Lucian Georgescu for setting up the ASR system. 

\bibliographystyle{IEEEtran}
\bibliography{master}

\end{document}